\documentclass{fac}
\usepackage[utf8]{inputenc}
\usepackage[hyphens]{url}
\usepackage{amssymb,graphicx}
\usepackage{isabelle,isabellesym}
\isabellestyle{literal}
\renewcommand{\isastyle}{\isastyleminor}

\renewenvironment{quote}
               {\list{}{\rightmargin\leftmargin}%
                \item\relax}
               {\endlist}

\title{From LCF to Isabelle/HOL}

\author[Lawrence C. Paulson, Tobias Nipkow and Makarius Wenzel]
    {Lawrence C. Paulson$^1$, Tobias Nipkow$^2$ and Makarius Wenzel$^3$\\
     $^1$ Computer Laboratory, University of Cambridge, UK\\
     $^2$ Fakult\"at f\"ur Informatik, Technische Universit\"at M\"unchen, Germany\\
     $^3$ Augsburg, Germany}

\correspond{Lawrence C. Paulson, Computer Laboratory, University of Cambridge, 15 JJ Thomson Avenue, Cambridge CB3 0FD, England.
            e-mail: lp15@cam.ac.uk}

\pubyear{2019}
\let\ts=\thinspace
\begin{document}

\makecorrespond

\maketitle

\begin{abstract}
Interactive theorem provers have developed dramatically over the past four decades, from primitive beginnings to today's powerful systems. Here, we focus on Isabelle/HOL and its distinctive strengths. They include automatic proof search, borrowing techniques from the world of first order theorem proving, but also the automatic search for counterexamples. They include a highly readable structured language of proofs and a unique interactive development environment for editing live proof documents. Everything rests on the foundation conceived by Robin Milner for Edinburgh LCF: a proof kernel, using abstract types to ensure soundness and eliminate the need to store proofs. Compared with the research prototypes of the 1970s, Isabelle is a practical and versatile tool. It is used by system designers, mathematicians and many others.
\end{abstract}

\begin{keywords}
LCF, HOL, Isabelle, interactive theorem proving.
\end{keywords}

\section{Introduction}

Today's interactive theorem provers originated in research undertaken during the 1970s on the verification of functional programs. Two quite different tools were built: the Boyer/Moore theorem prover (now ACL2, described elsewhere in this volume) and Edinburgh LCF \cite{mgordon79,paulson-computational}.

Descendants of the latter include every member of the HOL family (HOL4, HOL Light, ProofPower) \cite{GordonM-HOL} as well as Coq \cite{BertotC04} and Isabelle \cite{isa-tutorial}. As we shall see in the sequel, the achievements of the past 40 years lie on several dimensions, including type systems, proof languages and user interfaces. Automation is the key to usability; this includes automated search for counterexamples as well as for proofs.

These developments are reflected in the size of the tools themselves. In 1977, the Edinburgh LCF distribution was 900 KB, including an implementation of the ML programming language in Lisp. In 1986, the Isabelle distribution was a mere 324 KB (not counting ML, now a separate distribution). By 1991, Isabelle exceeded LCF: 937 KB\@. In 2019, the Isabelle distribution had reached 133 MB\@! And by way of comparison, HOL Light (which is much more closely related to LCF) was 84 MB\@. Much of this bulk consists of proof libraries rather than executable code, though libraries make a crucial contribution to a system's capabilities. Today's mature systems also include documentation and examples.%
\footnote{The figures are for uncompressed distribution directories containing no binaries, but possibly PDF files.}

Isabelle is a leading interactive theorem prover. It is generic, supporting a number of different formal calculi, but by far the most important of these is its instantiation to higher-order logic: Isabelle/HOL. Already during the 1990s, Isabelle/HOL was being applied with great success to the task of verifying cryptographic protocols \cite{paulson-jcs}. Turning to mathematics, it played a critical role in Hales's Flyspeck project, which verified his proof of the Kepler conjecture \cite{hales-formal-Kepler}. It is the basis for the seL4 project, under which an entire operating system kernel was verified, proving full functional correctness \cite{klein-sel4}. It was adopted by researchers outside the verification milieu for specifying and verifying algorithms for replicated datatypes that provide ``eventual consistency'' \cite{gomes-verifying-strong}. Numerous other projects are underway around the world. Like other proof assistants, Isabelle is not directly concerned with program verification, i.e. with verifying code written in a programming language, but it can be used as a back end to prove verification conditions. Isabelle can also be used as a verified programming environment, where mathematical functions can be proved correct and then automatically translated to executable code in one of several different programming languages. The translation process itself is currently unverified, but even this is likely to change in the near future.

This essay focuses on Isabelle/HOL and therefore has little to say about techniques common to most systems. For example, simplification by rewriting --- coupled with recursive simplification to handle conditional rewrite rules --- was already realised in both the Boyer/Moore theorem prover \cite{bm79} and Edinburgh LCF by the end of the 1970s. Recursive datatype and function definitions, as well as inductive definitions, were commonplace by around 1995. Linear arithmetic decision procedures were also widely available by then.

Our title echoes Mike Gordon's paper ``From LCF to HOL: a Short History'' \cite{mgordon-history}. Like Mike, we begin with LCF, the source of the most fundamental ideas. But we pass over this material quickly in order to focus on Isabelle. There is no way to surpass Mike's account of the early years. He starts in 1969, with Dana Scott's Logic for Computable Functions, and covers the original LCF project at Stanford University. He describes in detail the development of the successor LCF systems at Edinburgh and then at Cambridge. He names all the people involved and finally outlines his own development of HOL\@.

We begin with LCF because of its seminal importance, continuing to HOL because it is so strongly linked (\S\ref{sec:lcf-and-hol}). Then we focus exclusively on Isabelle. We begin with the core ideas of a generic reasoner built around unification and backtracking (\S\ref{sec:early}). Then we consider the task of supporting higher-order logic, which required the introduction of type classes (\S\ref{sec:type classes}). There follows an account of automatic proof search and its dual, the search for counterexamples (\S\ref{sec:automation}--\ref{sec:counterx}). We also discuss the generation of code from logical functions (\S\ref{sec:code}). We then turn to Isabelle's most distinctive features: its structured proof language (\S\ref{sec:isar}) and the powerful user interface architecture supporting it (\S\ref{sec:PIDE}). The next section describes Isabelle's Archive of Formal Proofs (\S\ref{sec:afp}). To conclude, we discuss the synergy among these ideas, making them more powerful in combination than individually (\S\ref{sec:post}).

Coq and other systems built around constructive type theories represent a distinctive strand of development and fall outside our scope.

\section{LCF and HOL}\label{sec:lcf-and-hol}

Edinburgh LCF is best known for the so-called \textit{LCF approach}: 
implementing the inference rules of a logical calculus within a \textit{proof kernel} that has the exclusive right to create theorems. 
Such a kernel is indeed found in most modern systems, and is responsible for their good record of soundness. But in fact, LCF introduced a broader set of norms that are now taken for granted: a focus on backward proof, the practice of working in a theory hierarchy, and above all, the central role of a functional programming language, ML \cite{milner85}.

ML is short for \emph{meta language}, as it was the vehicle for operating on
formulas belonging to the \emph{object language} (namely, the proof calculus).
Radical at the time, ML was soon seen to be a general-purpose programming
language and today exerts a strong influence on language design. 
Crucial to ML is its sound, polymorphic type system with first-class functions, 
and in particular, its support for abstract types. An \emph{abstract
type} encapsulates a data structure's internal representation, allowing access only through a
fixed set of functions. A typical abstract type might be a dictionary, where the
 implementation (e.g.\ balanced trees) is only accessible through well-defined
operations such as insert, update and delete; then the implementer is free to
change the representation (e.g.\ to improve performance or to introduce additional
operations) without affecting any of the code outside the abstract data type
declaration itself.

 Robin Milner's key insight was that abstract data types could save space by eliminating
 the need to store proofs. (In his early experiments, he had kept running out of
 memory \cite{gordon-tactics-milner}.) He declared an abstract type of theorems, where the operations were
 simply the inference rules of his logical calculus. The resulting type,
 \texttt{thm}, was the type of theorems and ML's type checker was our guarantee
 that anything of type \texttt{thm} had definitely been created exclusively by
 the application of inference rules. Just as a dictionary doesn't need to keep a
 record of the operations applied to it but only the dictionary itself, type
 \texttt{thm} doesn't need to store the proofs of theorems but only their
 statements. The resultant savings of space are as important now as they were in
 the 1970s, for although today we have more memory, we also have vastly bigger
 proofs.

Edinburgh LCF also introduced proof \emph{tactics} and \emph{tacticals} to
express backward, goal-directed reasoning. An LCF tactic typically applies a
specific inference rule, while tacticals denote control structures, e.g.,
\texttt{THEN} (one tactic followed by another), \texttt{ORELSE} (which allows
one of several tactics to be attempted) and \texttt{REPEAT} (repeated execution
of a tactic until it fails). Remarkably, even tactics lie outside the proof
kernel. An LCF tactic is a function that takes a goal (a formula with its
assumptions) and returns a list of subgoals that it claims are logically
sufficient. To justify this claim, it returns a function operating on values of
type \texttt{thm}, but we have no guarantee that this function will deliver the
promised theorem in the end. We see that the LCF architecture makes proof
procedures harder to implement, while reducing the amount of code that has to be
trusted.

We see incidentally two meanings of the word \emph{proof}:
\begin{enumerate}
  \item formal deductions of theorems from axioms using the inference rules of a logical calculus;
  \item executable code written using tactics or other primitives, expressing the search for such deductions.
\end{enumerate}
To resolve this ambiguity, the former are sometimes called \emph{proof objects} or \emph{proof terms} and the latter, 
\emph{proof scripts} or \emph{proof texts}.
Thus we see that the LCF approach eliminates the need to store proof terms and allows proof scripts to be coded in ML using tactics and tacticals. Nevertheless, proof assistants based on constructive type theories retain proof objects, as they are intrinsic to such formalisms.

The most fundamental question in the design of a theorem prover is what calculus to support. Boyer and Moore made the inspired choice of Pure Lisp, which was sufficient for verifying simple functional programs and formalising elementary number theory in the 1970s, and which has grown to support advanced applications today. For Edinburgh LCF, Milner chose the Logic for Computable Functions, which was perfect for the domain-theoretic investigations topical at the time but proved to be too quirky for general adoption.

Mike Gordon, who was one of the designers of Edinburgh LCF, ultimately adopted higher-order logic as the basis for his research into hardware verification~\cite{mgordon86,mgordon-history}. Higher-order logic is most easily understood as a typed predicate calculus including function and Boolean types, and therefore also set types. Gordon launched his HOL system in 1986~\cite{mgordon86,GordonM-HOL}, presenting both the logic and its application to hardware specification and verification. HOL turned out to be extremely versatile and it soon attracted a global user community. New versions emerged, in particular HOL Light \cite{hol-light-tutorial}. Members of the HOL family have been used in verification projects of every description, including the formalisation of great bodies of mathematics.

 However, it wasn't obvious in the 1980s that one or two formalisms could be sufficient for the needs of verification. Type theories, dynamic logics and many other formalisms were being proposed. Gordon himself had reworked LCF twice in his hardware verification research. This was the origin of the idea that a theorem prover could be \emph{generic}: supporting a spectrum of formalisms through a common framework. Today this includes syntactic tools (parsing, pretty printing), inference tools (rewriting, unification), a common proof language (Isar) and user interface foundation (PIDE).

\section{Isabelle in The Early Days: A Logical Framework}\label{sec:early}

Isabelle originated in a project to build an LCF-style proof assistant for Martin-L\"of's constructive type theory \cite{martinlof85}. Two ideas influenced the design from the outset \cite{paulson700}:

\begin{itemize}
  \item Reasoning should be based on \emph{unification} rather than pattern matching, so that goals could contain variables that could be instantiated.
\item A tactic should be able to return multiple results in a lazy list, representing alternative proof attempts for \emph{backtracking}.
\end{itemize}
Both ideas were aimed at supporting proof search. The LCF work had demonstrated 
conclusively that verification was repetitious and tedious, requiring the best possible automation. 
The combination of unification and backtracking would allow the use of logic programming techniques, as in Prolog \cite{clocksin-mellish}. For example, in Martin-L\"of's type theory we have the following (derived) rule:
\[ \frac{c\in A\times B}{\mathrm{fst}(c)\in A.} \]
Backward chaining, using this rule to prove something of the form $\mathrm{fst}(c)\in A$, 
leaves us with a subgoal of the form $c\in A\times {?B}$, where the question mark indicates that $?B$ is a hole  to be filled: in Prolog parlance, a logical variable. Unification fills in the holes later, while backtracking can orchestrate alternative ways of filling the holes. 
In this example, the form of $c$ will determine the choice of~$?B$.
The ability to defer choices until they become clear is as valuable in theorem proving as it is in Prolog programming. 

It may be worth mentioning that Prolog was a hot topic in the 1980s.
However, standard first-order unification was simply not applicable to
Martin-L\"of type theory, which is based on a form of typed $\lambda$-calculus.
\textit{Higher-order unification} would be necessary. Undecidable in its full generality,
a reasonably practical unification procedure had recently been published by
Gerard Huet \cite{huet75}. 

Higher-order unification is strikingly different from
first-order unification. The latter is easily implemented and delivers a unique
answer modulo variable renaming. Higher-order unification allows even functions
to be variables and can instantiate such variables with $\lambda$-abstractions
generated on the fly, yielding multiple results. For example, it is possible to
unify $F\, M$ with 3 in two different ways: $F=\lambda x.\, 3$ (with no
constraint on $M$) and $F=\lambda x.\, x$, $M=3$. Unifying $F\, 3$ with $3+3$ could
yield four different results: $F=\lambda x.\, x+x$,  $F=\lambda x.\, x+3$,
$F=\lambda x.\, 3+x$ and $F=\lambda x.\, 3+3$. The search for such unifiers
works by descending through the terms one level at a time, attempting two kinds
of steps: \emph{projections} (use of a bound variable) and \emph{imitations}
(copying the opposite term). The examples above show how it can find different
ways of abstracting one term from another.

With such a sophisticated syntactic mechanism built in at the lowest level, it became clear that the LCF approach to inference rules could be radically changed \cite{paulson-natural}. Consider the proof rule of universal elimination:
\begin{equation}
\frac{\forall x.\, \phi(x)}{\phi(a)}  \label{eqn:spec}
\end{equation}
Its LCF representation is an ML function taking two arguments: a theorem (which must have the form $\forall x.\, \phi(x)$) and a term $a$ (which must have the same type as~$x$). It then generates the desired conclusion, namely $\phi(a)$, raising an exception if any of the preconditions is violated. This approach is general; the drawback is the tedium and attendant risk of error when there are dozens of rules. With our $\lambda$-calculus framework, inference rules such as the one above can simply be written out in the form of a template and instantiated using unification. Better still, unification could be applied either to the premise or to the conclusion, yielding forward or backward proof through a single mechanism. This is Isabelle's central idea \cite{paulson-natural}.

The precise nature of these templates remained to be determined, and the best approach turned out to be a sort of \emph{logical framework}. These are specialised formalisms whose purpose is to encode other formalisms. The Edinburgh Logical Framework \cite{harper-jacm} is the best known of these. It differs from Isabelle's by incorporating proof objects into the calculus itself, neutralising one of the key advantages of the LCF architecture: that proofs do not have to be stored.

 So Isabelle can be seen as an instance of the traditional LCF approach, but where type \texttt{thm} formalises a logical framework or \textit{meta-logic}, where other formalisms (the \textit{object-logics}) can later be encoded. Such encodings can also be proved correct \cite{paulson-found}.
Our logical framework approach is less general than the original LCF representation, where an inference rule can undertake an arbitrary computation. Nevertheless, it captures a variety of possibilities, and moreover, it is open-ended: for example, Isabelle defines intuitionistic first-order logic, which in a succession of formal theories is extended with classical logic, then with Zermelo--Fraenkel set theory, then with the axiom of choice. With the original LCF approach, once you define the type \texttt{thm}, it can never be extended.

The first object-logic was constructive type theory (Isabelle/CTT). It was followed by a classical first-order sequent calculus (Isabelle/LK) and by natural deduction calculi for intuitionistic and classical first-order logic (Isabelle/IFOL and Isabelle/FOL) \cite{isabelle-logics}. Later, a substantial development of Zermelo--Fraenkel set theory was developed on top of Isabelle/FOL \cite{paulson-gr}. This is one of the leading tools for formal reasoning in axiomatic set theory.

Although Isabelle took a declarative approach to defining logics, the system
architecture was still based on LCF's philosophy that everything was done in
ML\@. All interactions with Isabelle took place at the ML toplevel. For example, the declaration of new types, constants with their definitions, and axioms all required calling appropriate ML functions with any necessary data as arguments. This began to change
in the early 1990s \cite{Paulson94} when logics could be defined in a \emph{theory file} with types, constants and proof rules declared using a natural syntax.\footnote{This represented a return to the original Edinburgh LCF, which also supported theory files. The Isabelle version was inspired by Goguen's OBJ system
\cite{Goguen78,FutatsugiGJ-POPL85}, in particular concerning the declaration of new mixfix syntax.}
Proofs, however, were still expressed in ML\@.  The structured proof language
Isar (Section~\ref{sec:Isar}) came later. Of today's LCF-based systems, the HOL family has remained the most faithful to the original conception, with proofs coded in ML\@.

Paulson's original implementation of higher-order unification~\cite{paulson-natural} was practical, but still slow.
Yet in Isabelle practice, many unification problems are
first-order, or almost so. Dale Miller \cite{Miller-JLC91} discovered
a subclass of $\lambda$-terms, later called (higher-order) patterns
\cite{Nipkow-LICS91}, which behave like first-order terms:
unification is decidable and if two terms are unifiable, they have a
\emph{most-general unifier}. A term (in $\beta$-normal form) is called a
\emph{higher-order pattern} if every free occurrence of a variable has as arguments
a list of distinct bound variables. Most unification problems in Isabelle are
of this form. Nipkow gave a succinct implementation of pattern
unification and added it to Isabelle~\cite{Nipkow-LICS93}. Full higher-order
unification is invoked only if pattern unification encounters a non-pattern.

A special case of unification is matching where the variables of only one of
the two terms are instantiated.  Isabelle's rewrite engine (aka the
\emph{simplifier}) is based on higher-order pattern matching.  Thus the
simplifier can deal with many standard transformations of quantified terms,
for example the following ones:
\[
\begin{array}{rcl}
(\forall x.\, P(x) \land Q(x)) &=& (\forall x.\, P(x)) \land (\forall x.\, Q(x)) \\
(\forall x.\, P \lor Q(x)) &=& P \lor (\forall x.\, Q(x)) \\
(\forall x.\, x = t \land P(x)) &=& P(t)
\end{array}
\]
It appears that Isabelle was the first theorem prover to support
higher-order rewrite rules \cite{NipkowP98}.

\section{Type Classes and Isabelle/HOL}\label{sec:type classes}

Gordon's HOL system became a runaway success, dominating the verification arena. 
By 1991 it was being used in over 80 separate projects around the world \cite{kalvala-hol}.
Isabelle claimed to be a generic theorem prover, but it couldn't handle higher-order logic. The problem was that the templates mentioned above could not refer to types. This was not an issue in the original application of Martin-L\"of type theory, where types and formulas were effectively identified and the inference rules referred to types explicitly. In the following example, $A$ and $A+B$ are types and the rule expresses how type checking should be done for a term of the form $\mathrm{inl}(a)$.
\[ \frac{a\in A}{\mathrm{inl}(a)\in A+B} \]
Contrast with the previous inference rule~(\ref{eqn:spec}), where no types are visible. There are many formalisms, including many-sorted first-order logic%
\footnote{Beware of terminological confusion regarding the word \textit{sort}, which for first-order logic is synonymous with \textit{type}. Our use of \textit{sort} below will be entirely different.}
as well as higher-order logic, where types are kept in the background and type constraints are enforced implicitly.
Users would not like to be forced to prove statements like $i + 1 : \mathrm{int}$.

Adequate support for logics having implicitly-typed variables required another idea, order-sorted polymorphism, leading to an even more powerful idea, axiomatic type classes.

\subsection{Order-Sorted Polymorphism}

Isabelle had always supported polymorphism internally; the difficulty lay in making it available to users. In its simplest form, a polymorphic type variable may take on any type whatsoever. This could not be allowed in a logical framework, where some types are intrinsic to the framework itself and other types might be unsuitable to a particular object-logic. In first-order logic, the type of $x$ in $\forall x.\, \phi(x)$ must not involve functions or Booleans.\footnote{Quantification over Booleans (and therefore, relations) requires at least second-order logic \cite[page~7]{barwise-fol}.}
\emph{Order-sorted polymorphism} solves such difficulties by introducing a hierarchy of sorts on types~\cite{Nipkow-CTRS-90}. A \emph{sort} is a finite intersection of \emph{type classes}, which are essentially collections of types. For first-order logic, we may introduce the type class~\isa{FO} for all types for which quantification is permitted. For higher order logic, we would have a different type class (say~\isa{HO}), containing Boolean and function types. We write $\tau :: C$
if type $\tau$ has class $C$.

Now we need to express how (possibly nullary) \emph{type constructors} such as \isa{nat} (the type of natural numbers), Cartesian product, function space and \isa{list} (the type of lists) act on type classes. For example, for first-order logic we want that \isa{nat {\isacharcolon\isacharcolon} FO} and that if $\tau$ \isa{{\isacharcolon\isacharcolon} FO} then $\tau$~\isa{list {\isacharcolon\isacharcolon} FO}. These assertions can be codified in Isabelle as instance declarations:
\begin{isabelle}
\ \ \isakeyword{instance} nat {\isacharcolon\isacharcolon} FO\isanewline
\ \ \isakeyword{instance} list {\isacharcolon\isacharcolon} {\isacharparenleft}FO{\isacharparenright} FO
\end{isabelle}
Thus \isa{nat list {\isacharcolon\isacharcolon} FO} holds, but in the absence of corresponding instances, 
\isa{bool {\isacharcolon\isacharcolon} FO} and \isa{bool list {\isacharcolon\isacharcolon} FO} do not:
truth values, or lists of them, do not belong to~\isa{FO}\@.
Technically, these instance declarations form the \emph{signature} of the (term) algebra of types.
Under certain natural conditions on the interaction of the class hierachy with the instance declarations, order-sorted unification of types, like unsorted unification, is still \emph{unitary}: solvable unification problems have \emph{most-general unifiers}. Therefore these conditions guarantee that we still have \emph{principal types}.
This theory, combined with extensions to higher-order unification~\cite{Nipkow-CTRS-90}, allows full support for higher-order logic. And so the ``templates'' of our framework contain explicit type variables, in addition to the ordinary variables that we had before. This version was first released in 1991~\cite{NipkowP92}. At the same time, the correspondence to Haskell's type classes was worked out~\cite{Nipkow-Snelting,Nipkow-Prehofer}.

\subsection{Axiomatic Type Classes}

The full power of type classes is realised when they are combined with axioms. \emph{Axiomatic type classes} support the flexible and principled overloading of symbols \cite{Nipkow-LF-91,wenzel-type}. A type class can be introduced on the basis of a specific vocabulary of symbols (or \emph{signature}) possibly coupled with axioms (a \emph{specification}) to constrain those symbols---and possibly extending other type classes. Orderings are a natural example: Isabelle/HOL introduces a succession of type classes for increasingly stronger notions of ordering:
\begin{itemize}
	\item \isa{ord}: the ordering symbols $<$, $\le$, but with no attached properties
	\item \isa{preorder}: adding reflexivity and transitivity, and defining $x < y$ iff $x \le y \land \neg (y \le x)$
	\item \isa{order}: adding antisymmetry
	\item \isa{linorder}: adding linearity
\end{itemize}
For the case of lists, the lexicographic ordering yields a partial ordering if the list elements are partially ordered and yields a linear ordering if the list elements are linearly ordered. Such details are easily expressible:
\begin{isabelle}
\ \ \isakeyword{instance} list {\isacharcolon\isacharcolon} {\isacharparenleft}order{\isacharparenright} order\isanewline
\ \ \isakeyword{instance} list {\isacharcolon\isacharcolon} {\isacharparenleft}linorder{\isacharparenright} linorder
\end{isabelle}
Instance declarations for axiomatic type classes require the user to supply definitions of any associated symbols (here $<$ and $\le$) for the supplied type along with proofs of the associated properties using those definitions. This overloading is principled in that although $<$ and $\le$ will have separate definitions for each type for which they are defined, they will always satisfy the associated axioms.

The value of axiomatic type classes can be seen in the formalisation of mathematical analysis. Type classes for groups, rings and other algebraic structures provide overloading for the common arithmetic symbols, but in addition, we have type classes for metric and topological spaces, vector and Euclidean spaces. When we construct the type of complex numbers for example, by showing that they form a field and a Euclidean space, we instantly inherit substantial libraries of facts covering limits, convergence, derivatives and topology, which would otherwise have to be largely duplicated from the analogous facts for the real numbers \cite{hoelzl-filters}. The same thing happens again when we construct more advanced number systems, such as the quaternions \cite{wood-quaternions}, a number system for three-dimensional space. 
We have a surprising variety of numeric types, such as the extended reals ($\mathbb{R}\cup\{+\infty,-\infty\}$) 
and the nonnegative extended reals; both of these belong to a number of ordering and topological type classes.

An exposition of axiomatic type classes with fully worked out examples is available in a paper by Paulson \cite{paulson-numerical}. It's slightly outdated --- we have a much more elaborate type class hierarchy now and a somewhat different syntax --- but the principles are the same.

\subsection{Logical Foundations}

Wenzel~\cite{wenzel-type} reduced type classes to (1) overloaded constant
definitions (OCDs) for the signature and (2) predicate definitions over a
single type variable for the specification. Getting the notion of
definition with type-polymorphism right is notoriously tricky.
For example, in an early version of Gordon's HOL system, constant
definitions could introduce inconsistencies \cite{Arthan16}. The problem
is simply to detect
ill-formed definitions, in particular circularities, automatically.
Circularities can lead to inconsistencies.
Type classes link the level of types with the level of terms,
opening up new opportunities to create highly obscure circularities.

Wenzel analysed the impact of OCDs on the consistency of an arbitrary
theory. He sketched conditions under which they were \emph{meta-safe}, which
roughly means that they can be removed without affecting
provability. Meta-safety implies conservativity, which in turn implies
consistency preservation.  However, his proof sketch considered an idealized
version of Isabelle's actual definition facilities, which were only partially implemented and
in conflict with existing application theories. In particular---in order to stay within Gordon's HOL
and the Isabelle logical framework---he excluded
the interleaving of OCDs with type definitions.  

Obua~\cite{Obua06}
found that Isabelle accepted OCDs that introduced inconsistencies.  He gave a
more rigorous and general formulation of Wenzel's
conditions~\cite{wenzel-type} and implemented them using an external
termination checker, outside of the inference kernel. He also sketched a proof that these conditions ensured
conservativity.  As an alternative, Wenzel and Haftmann proposed a much simpler (and
stronger) check on OCDs~\cite{HaftmannW06} inside the logical kernel, but this excluded a few application
theories with ambitious overloading.  

Several years later
Kun\v{c}ar and Popescu~\cite{KuncarP18,KuncarP-POPL18} rediscovered that combining
overloaded constant definitions with type definitions could
introduce circularities that were missed by the previous analyses.  They
gave extended checks to avoid these circularities and proved that these
checks ensure consistency of theories that extend the initial HOL theory
(comprising the standard HOL axioms) with overloaded constant
definitions and type definitions.  In the
end~\cite{KuncarP-POPL18}, they strengthened the result to show that such
definitional extensions are in fact meta-safe over the initial HOL theory.

These particular concerns were specific to detecting circular definitions, but all proof assistants potentially contain errors, like any other software. A notable case is PVS~\cite{Owre06}, which lacking an LCF-style proof kernel was particularly vulnerable to soundness errors (at least in the 1990s):

\begin{quote}
PVS still seems to contain a lot of bugs and frequently new bugs show up. \ldots It would be desirable that the bugs in PVS would only influence completeness and not soundness. Unfortunately, this is not the case, as some recent proofs of \texttt{true==false} have shown [15, bug numbers 71, 82, 113 and 160]. \ldots It is reasonable to assume that PVS will continue to contain soundness bugs. The obvious question thus arises, why use a proof tool that probably contains soundness bugs? Our answer is threefold: PVS is still a very critical reader of proofs. PVS lets fewer mistakes slip through than many of our human colleagues\ldots.
Furthermore, history tells us that the fixed soundness bugs are hardly ever unintentionally explored, we know of only a single case.
Thirdly, most mistakes in a system that is to be verified are detected in the process of making a formal specification.\cite[page 134--5]{griffioen-comparison-pvs}
\end{quote}

Notwithstanding such forbearance, users deserve a much higher standard of correctness than this. At the same time, it's vital to stress that users need to take responsibility for their own definitions: we know how to detect circularity, but no system can check whether a definition conforms to the user's true intentions.

\subsection{Isabelle/HOL versus HOL}

As mentioned at the start of this section, the purpose of the foregoing work was to make possible an Isabelle instantiation of higher-order logic. Isabelle/HOL emerged in 1991 \cite{NipkowP92} and was soon a fully capable alternative to HOL \cite{isa-tutorial}. In some ways it emulated the latter, particularly in its axiom system. Nevertheless, it is thoroughly Isabelle in its treatment of theorems, tactics, etc; we cannot regard Isabelle/HOL as a member of the HOL family the way that HOL Light is. 

Both Isabelle and HOL implement Milner's idea of a proof kernel implementing a formal calculus as an abstract type called~\texttt{thm}. The essential difference is that Isabelle's formal calculus is a logical framework in which other formalisms can be defined, while HOL's is simply higher-order logic. Either way, a function of type \texttt{thm->thm} implements an inference rule. In the case of Isabelle, this would be an operation at the level of the logical framework, such as the resolution rule through which proofs are constructed. In the case of HOL, this would be an inference in higher-order logic itself, like the quantifier rule~(\ref{eqn:spec}). In HOL, a tactic must be coded separately corresponding to each rule of inference, while in Isabelle, inference rules such as~(\ref{eqn:spec}) are all expressed declaratively and applied using the generic resolution rule.

Isabelle's generic nature can be seen in how many of its capabilities are shared among its various instances. The common libraries include a general representation of syntax, with parsing, pretty printing, type checking, simplification and other proof procedures, as well as the proof language and user interface. They are available in other instances of Isabelle, such as Isabelle/ZF, which has been the basis for substantial developments in Zermelo--Fraenkel set theory \cite{paulson-consistency}.

\section{Automation} \label{sec:automation}

It was clear from the outset that machine proof was extremely laborious and could only be feasible if the machine itself provided as much automation as possible. But it was also clear that fully automatic theorem proving was not practically achievable. Significant advances in automation had already been made by the end of the 1970s:

\begin{enumerate}
  \item decision procedures for arithmetic, arrays, lists, etc., as well as methods for combining decision procedures \cite{nelson-fast};
\item resolution for first-order logic \cite{robinson65,overbeek-hyper}, complete in principle but frequently disappointing in practice;
\item the signature automation of the Boyer/Moore theorem prover \cite{bm79}: conditional rewriting plus powerful heuristics for induction.
\end{enumerate}
Of these, simplifiers based on rewriting (by previously proved theorems of the form $t=u$) quickly found their way into LCF, HOL, etc\@. All such simplifiers eventually supported \textit{conditional rewriting}, for rewrites of the form $\phi \Longrightarrow t=u$; in such a case, the necessary instance of $\phi$ would be proved recursively by the simplifier itself. Arithmetic decision procedures were eventually adopted in many systems. On the other hand, resolution had acquired a bad reputation. As late as 2002, Shankar could write

\begin{quote}
The popularity of uniform proof methods like resolution stems from the simple dogma that since first-order logic is a generic language for expressing statements, generic first-order proof search methods must also be adequate for finding proofs. This central dogma seems absurd on the face of it. \ldots A more sophisticated version
of the dogma is that a uniform proof method can serve as the basic structure for introducing domain-specific automation. There is little empirical evidence that even this dogma has any validity.
\ldots Automated reasoning has for too long been identified with uniform proof search procedures in first-order logic. This approach shows very little promise.	\cite[pages 3--4]{shankar-little}
\end{quote}

Of the various proof assistants, only Isabelle saw a sustained effort to incorporate ideas from resolution.

\subsection{The classical reasoner}\label{sec:classical}

As mentioned in Section~\ref{sec:early} above, Isabelle supported both unification and backtracking from the start, with the aim of incorporating ideas from first-order automatic proof procedures. In the context of interactive proof, unification provided the ability to prove a subgoal of the form $\exists x.\, \phi(x)$ by removing the quantifier and proving~$\phi(?t)$, where $?t$ stood as a placeholder for a concrete term to be supplied later. Through unification, this term could even be built up incrementally. Dually, unification provided a means of using a universally quantified fact $\forall x.\, \phi(x)$, when the required instances were not immediately obvious.

Simple automation is achievable through a combination of obvious applications of the propositional connectives ($\land$, $\lor$, $\neg$, etc.) along with heuristics for performing quantifier reasoning. Stronger automation is obtainable by borrowing well-known techniques for  classical first-order logic theorem proving. But the most important idea is to embrace the concepts of natural deduction in application theories as well as in pure logic. Natural deduction prefers the use of simple inference rules focusing on a single symbol. For example, conjunction is effectively defined by the following three rules:
\[ \frac{\phi \quad \psi}{\phi\land \psi} \qquad \frac{\phi\land \psi}{\phi}\qquad \frac{\phi\land \psi}{\psi}\]
The intersection of two sets has a technical definition that would greatly complicate reasoning, but it is easy to derive inference rules for intersection in the style of natural deduction (and analogous to those above):
\[ \frac{a\in A \quad a\in B}{a\in A\cap B} \qquad \frac{a\in A\cap B}{a\in A}\qquad \frac{a\in A\cap B}{a\in B}\]
Many other reasoning steps can be expressed similarly:
\[ \frac{A\subseteq B \quad a\in A}{a\in B} \qquad
   \frac{A\subseteq B \quad B\subseteq A}{A=B} \qquad
   \frac{k \mathrel{\mathbf{dvd}} m\quad k \mathrel{\mathbf{dvd}} n}{k \mathrel{\mathbf{dvd}} \gcd(m,n)}\]
So the crucial idea is to build proof tactics that support reasoning with inference rules of this sort. Though they borrow techniques from first-logic theorem provers, they are far more effective than expanding the various operators into their low-level definitions and attempting to prove the resulting formulas of pure logic. And note: they would be formulas of higher-order logic, where automation is considerably more difficult than for first-order logic.

Such tactics have collectively become known as Isabelle's \emph{classical reasoner} \cite{paulson-blast}. A principle of natural deduction is that the syntactic form of each rule clearly identifies which symbol it is concerned with. And therefore hundreds of such rules can coexist in the classical reasoner without causing a combinatorial explosion. The user who invokes \isa{auto}---which combines classical reasoning and simplification---gains the benefit of built-in knowledge about everything in Isabelle's standard libraries. As users build their own libraries, they can continue to augment this knowledge. This combination of classical reasoning with rewriting is still unique to Isabelle.

\subsection{Sledgehammer}

By the early 2000s, resolution theorem provers \cite{vampire02,schulz-e,weidenbach-spass} were demonstrating power far beyond anything the classical reasoner could ever achieve. The idea of some sort of interface between Isabelle and these systems was beguiling. This idea wasn't new: such linkups have been attempted on a number of past occasions, always unsuccessfully. The key was to make such a linkup useful:
\par
\begin{quote}
The guiding idea is that user interaction should be minimal. The system should invoke automatic provers spontaneously or in response to a trivial gesture such as a mouse click. These proof attempts should run in the background, not disturbing the user unless a proof is found. Proofs should refer to a large library of known lemmas: users should not have to select the relevant ones. The automatic prover should not be trusted; instead, proofs should be translated back into the formalism of the interactive prover. Proofs should be delivered in source form to the user, who can simply paste them into her proof script. \cite[p.\ts1576]{meng-automation-interactive}
\end{quote}
A major difficulty with building an interface between Isabelle/HOL and first-order automatic theorem provers is that they operate on quite different formalisms. Isabelle/HOL has $\lambda$-abstractions, types and type classes, while first-order logic has none of these. One click invocation could only be achieved if the system itself took care of everything: the translation of higher-order syntax, some representation of type constraints, etc. An additional requirement was \textit{relevance filtering}: to identify the most suitable of the thousands of facts  available in an Isabelle session, since providing too many would overwhelm the first-order provers.

The final difficulty was to translate the proofs discovered by the external reasoning tools into Isabelle's proof kernel. This translation moreover had to be expressed as a source-level proof so that the expensive proof search would not have to be repeated. Resolution theorem provers typically print a formal justification of their reasoning, but it is difficult to interpret and frequently ambiguous. The approach eventually adopted was to extract from their output nothing but the list of axioms---each a known Isabelle theorem---used in the proof. This list would typically contain no more than eight axioms, yielding a problem simple enough to be proved by simple tactics integrated with Isabelle's kernel \cite{paulson-susanto}. This tool became known as \emph{Sledgehammer}. The original version was developed at Cambridge, but Sledgehammer was comprehensively rewritten at Munich, particularly by Blanchette \cite{blanchette-extending}, who greatly increased its scope and power. 

\begin{figure}
  \centering
  \includegraphics[width=0.75\textwidth]{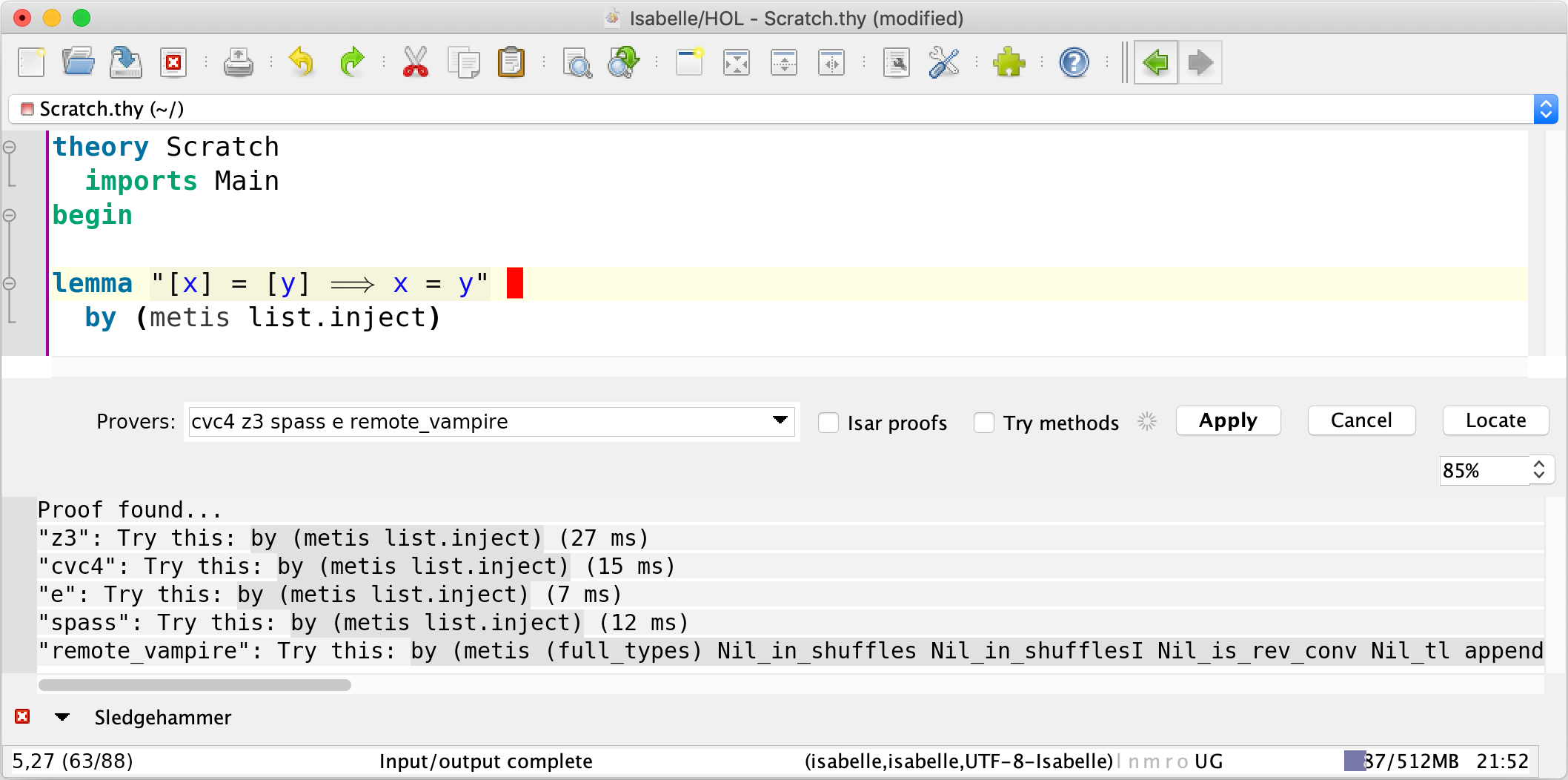}
  \caption{Sledgehammer GUI in Isabelle/jEdit: clicking on highlighted
    output inserts the proposed proof snippet into the text}
  \label{fig:sledgehammer}
\end{figure}

Sledgehammer helps beginners by identifying and using theorems that they didn't know existed. It puts the latest theorem provers, such as Vampire and Z3 \cite{moura-z3}, at their disposal with a single click (Figure~\ref{fig:sledgehammer}). But even advanced users are often surprised by the proofs Sledgehammer comes up with. It is now seen as indispensable, and similar subsystems are being developed for other interactive theorem provers.

\section{Counterexample Search} \label{sec:counterx}

Isabelle's proof methods and Sledgehammer are effective for proving \emph{theorems}, 
but given an invalid conjecture they normally fail to detect the
invalidity, let alone produce an informative counterexample. Novices and
experts alike state invalid theorems and find themselves wasting hours on 
impossible proofs. To make proving more enjoyable and productive, Isabelle
includes counterexample generators that complement the proof tools. The main
ones are Quickcheck and Nitpick. As a simple example, suppose that the user types in
this lemma statement (where \isa{rev} reverses a list and \isa{\isacharat}
appends two lists):

\begin{isabelle}
\ \ \isacommand{lemma}\isamarkupfalse%
\ {\isachardoublequoteopen}rev\ {\isacharparenleft}xs\ {\isacharat}\ ys{\isacharparenright}\ {\isacharequal}\ rev\ xs\ {\isacharat}\ rev\ ys{\isachardoublequoteclose}
\end{isabelle}
Quickcheck is invoked automatically and displays the counterexample
\isa{xs = {\isacharbrackleft}a$_1${\isacharbrackright}, ys = {\isacharbrackleft}a$_2${\isacharbrackright}}.

\subsection{Quickcheck}
\label{sec:quickcheck}

As the name suggests, Quickcheck is Isabelle's counterpart to the
QuickCheck testing tool~\cite{ClaessenH00} for Haskell.  The original
Quickcheck~\cite{Berghofer-Nipkow-SEFM2004} combined Isabelle's code
generation infrastructure (Section~\ref{sec:code}) with random testing,
covering both recursive functions and inductive predicates.  
It aimed at providing fully automatic counterexample search, in contrast to
Haskell's QuickCheck, which is an infrastructure for building specialized random generators for testing. 
Therefore it worked automatically whenever all functions and inductive
predicates were executable, but, depending on the property, it might take
forever to find a counterexample. Consider especially conditional
properties $\phi \Longrightarrow \psi$ where most of the random values simply
falsify $\phi$; this situation requires a special random generator that yields
only values that satisfy $\phi$.  

Therefore Bulwahn~\cite{Bulwahn-PhD,Bulwahn-CPP12,Bulwahn-LPAR12,BlanchetteBN-FROCOS11}
refined Quickcheck in three directions:
\begin{itemize}
\item Exhaustive enumeration of small values, an idea due to
Runciman {et al.} \cite{RuncimanNL08}.
\item Generating values not randomly but synthesizing generators from
the premises of conditional properties.
\item Symbolic testing with \emph{narrowing}, a technique from
\emph{functional-logic programming}~\cite{AntoyH10}.
\end{itemize}

Isabelle's Quickcheck was inspired by the work of Dybjer {et al.}
\cite{DybjerHT-TPHOLs03} in the theorem prover Agda: they had followed the
original QuickCheck design, which expected users to set up specialized random
generators in the host language/logic. Over the next decade, other major
theorem provers adopted analogous checkers: PVS~\cite{Owre06},
ACL2~\cite{ChamarthiDKM2011} and Coq~\cite{Paraskevopoulou15}.

Quickcheck is one of Isabelle's best-loved tools (after Sledgehammer), partly because it is
invoked automatically and silently every time the user types in a lemma. You
may not even be aware of its existence and it suddenly announces that it has
found a counterexample to your purported lemma. The result is a surprised and
grateful user. Quickcheck is most effective in the context of
functional programming combined with inductive predicates.

\subsection{Nitpick}

A radically different approach to Quickcheck is based on systematic model
enumeration using a SAT solver. This approach was pioneered by the tool
Refute~\cite{Weber05,Weber-PhD} and is now embodied by
Nitpick~\cite{BlanchetteN10,Blanchette-PhD}. Nitpick looks for finite
fragments (substructures) of infinite countermodels, soundly approximating
problematic constructs. Common Isabelle idioms, such as inductive and
coinductive predicates and datatypes as well as recursive and corecursive
functions, are treated specially to ensure efficient SAT solving
\cite{Blanchette-SQJ13}. The actual reduction to SAT is performed by the
Kodkod library \cite{TorlakJ07}.  Given a conjecture, Nitpick (via Kodkod and
the SAT solver) searches for a standard set-theoretic model that falsifies it
while satisfying any relevant axioms and definitions.  Nitpick is innately
better suited to problems from set theory and logic than
Quickcheck. Nitpick revels in particular in finite combinatorial problems.

The first tool that exploits SAT-solving for
finding counterexamples in a theorem prover (ACL2) seems to be due to
Sumners~\cite{Sumners02}. A second prototype tool, again for ACL2, was
developed by Spiridonov and Khurshid \cite{SpiridonovK07} and was based on
Kodkod. Blanchette was unaware of this work when developing Nitpick.

\section{Code Generation} \label{sec:code}

Code generation is the process of generating efficiently executable code from
definitions in the logic of a theorem prover. It serves three
main purposes: to obtain actual runnable software, to validate definitions by
executing them on concrete values, and to search for counterexamples to
properties by testing.

Most major theorem provers are based on logics that have an executable
sublanguage. In the Boyer and Moore theorem prover, from its earliest incarnation \cite{bm79}
to present-day ACL2 \cite{hunt-industrial}, the logic is itself a purely functional fragment of Lisp:
all expressions can be executed according to Lisp semantics.
In Martin-L\"of type theory \cite{martinlof85}, the term language 
--- including proof terms --- has a well-defined operational semantics.
Terms are also executable, subject to certain conditions, 
in other type-theory based systems such as Coq \cite{BertotC04}.
Note that the types (which correspond to the formulas of 
predicate logic) are not executable.

In the case of HOL, a sublanguage can be identified that
corresponds to a functional programming language.  
Then we superimpose an operational semantics on this fragment: 
programs are sets of equations that are to be used as rewrite rules, from left to
right. This idea goes back to term rewriting and was expressed succinctly by
titles like \emph{Computing in Systems Described by Equations}
\cite{ODonnell-LNCS} and \emph{Programming with Equations}
\cite{HoffmannOD-TOPLAS82}. The step-by-step execution of an equational
program by rewriting corresponds to performing a proof in equational logic.
If the rewrite rules come from recursive function definitions in the first
place, it is natural to translate them to programs in an ML-like functional language
with pattern-matching. The first such tool was created by
Rajan~\cite{Rajan92} for Gordon's HOL system~\cite{GordonM-HOL}.

\subsection{History}

Berghofer and Nipkow \cite{BerghoferN-TYPES00} realised the approach above, 
creating a compiler in Isabelle from a subset of HOL into ML\@. Datatypes and recursive
functions defined by pattern matching are translated directly into their ML
counterparts. They also extended the approach by making a subset of
inductive predicates executable. The idea is to view them as Prolog programs
and to perform a \emph{mode analysis} \cite{DebrayW88}. Each possible mode partitions the
arguments of an inductive predicate into inputs and outputs, and for each mode the inductive predicate is
compiled into an ML function that maps an input tuple to a stream of output
tuples. A subset of HOL terms and certain queries involving inductive
predicates can now be compiled and executed in ML\@. This formed the basis of
Isabelle's initial Quickcheck (Section~\ref{sec:quickcheck}).

Isabelle's code generator is part of the trusted kernel.  Work by
Haftmann \cite{Haftmann-PhD} and others improved its reliability and
functionality significantly: code generation for inductive predicates was
moved out of the kernel by translating inductive predicates into recursive
functions inside HOL, where the equivalence is proved
\cite{BerghoferBH09}. Reliability of the compilation of the purely functional
sublanguage was improved by a pen-and-paper correctness proof
\cite{HaftmannN-FLOPS2010}. Code generation was extended to type classes by
eliminating them in a first step. Code generation was extended to support
data refinement \cite{HaftmannKKN-ITP13}, i.e.\ the automatic implementation of abstract types like sets by concrete types like search trees.  Due to a new modular design,
further target languages are easier to add: as of this writing, the code
generator supports Standard~ML, OCaml, Haskell and Scala.

Most recently, Hupel \cite{HupelN-ESOP18} has provided an alternative
verified code generator that translates HOL into CakeML, an ML-like
functional language with a verified compiler~\cite{KumarMNO14}. 
CakeML is the only backend for code generation that doesn't have to be trusted, since the CakeML compiler has itself been verified formally.
This yields a verified tool chain from HOL to machine code, except that the CakeML compiler
was verified in HOL4 \cite{slind-hol4} rather than Isabelle/HOL\@. Eliminating this
gap is the subject of current research.

\subsection{Applications}

The code generator has been an enabling technology for a large number of applications. The following are some representative examples:
\begin{description}
\item[Imperative HOL]
Bulwahn {et al.}~\cite{BulwahnKHEM08}
developed a monadic embedding of imperative programs in HOL. That is, on the HOL level everything is still purely functional, but in a monadic style.
They extended the code generator such that it translates these monadic
definitions into actual imperative code in the target language (SML etc.).
This leads to substantially improved performance and is used in a number of applications below.

\item[Refinement Framework]
Lammich~\cite{Lammich13,cpp/Lammich16,jar/Lammich19} has developed an HOL framework for the stepwise refinement of (possibly nondeterministic) algorithms down to (possibly imperative) executable code. This framework is used in a number of applications below. Lammich and others \cite{itp/LammichT12,itp/Lammich14,jar/LammichS19} have used the framework extensively for the verification of efficient graph algorithms.

\item[Model Checking and SAT Solvers]
Esparza {et al.}~\cite{cav/EsparzaLNNSS13} developed an executable verified model LTL model checker (10--50 times slower than SPIN~\cite{DBLP:journals/tse/Holzmann97} on standard benchmarks) that was later extended with partial order reduction~\cite{jar/BrunnerL18}.
Brunner and Lammich built on work by Peled \cite{fmsd/Peled96} but found that one of his lemmas was incorrect; thus they were unable to use his actual reduction algorithm. Siegel~\cite{Siegel-CAV19} found a counterexample to the correctness of Peled's algorithm with the help of the Alloy analyzer~\cite{Jackson06}.

Wimmer and Lammich~\cite{tacas/WimmerL18} developed an executable model checker
for timed automata whose throughput is about one order of magnitude lower than Uppaal's~\cite{LarsenPY97} on standard benchmarks (but degenerates for large state spaces).

Lammich~\cite{cade/Lammich17} verified an executable checker for unsatisfiability certificates emitted by SAT solvers which is twice as fast as the standard unverified checker.

\item[Term Rewriting]
Thiemann has been developing a huge formalisation of the theory of term rewriting
called \emph{IsaFoR/CeTA}%
\footnote{\url{http://cl-informatik.uibk.ac.at/software/ceta/}}
over more than a decade
now \cite{ThiemannS09}. Initially, IsaFoR/CeTA was aimed primarily at checking
termination proofs found by automatic tools like AProVE~\cite{GieslABEFFHOPSS17}.
The code generator produces these proof checkers from their verified HOL formalizations. Today, IsaFoR/CeTA can also check proofs of
a term rewriting system's complexity~\cite{DivasonJKT018} and confluence~\cite{NageleM16}.

\item[Computer Algebra]
As representative examples we mention decision procedures for univariate
real polynomials (based on Sturm~\cite{eberl15cpp} and on cylindrical algebraic decomposition~\cite{jar/LiPP19})
and the Berlekamp-Zassenhaus factorization algorithm~\cite{jar/DivasonJTY19}.

\item[Programming Languages]
One key application of theorem provers has been the formalization of programming languages and compilers \cite{Leroy-JAR09,KumarMNO14,Concrete}. The code generator was used by Lochbihler and Bulwahn \cite{DBLP:conf/itp/LochbihlerB11} to generate an interpreter directly from the semantics of a Java-like language with threads.
\end{description}

\section{Structured Proofs, Structured Specifications and Formal Contexts} \label{sec:isar}

In the beginning, formal proofs were tiny and the ML proof scripts were readable enough. 
A typical proof goal involved just a few assumptions involving two or three bound variables. 
But as the field progressed, researchers tackled increasingly ambitious problems.
The longer proofs got, the more incomprehensible they became.
Moreover, the traditional LCF approach of tactical proof had a tendency to retain too much, 
so that the user might be faced with a list of several dozen assumptions. 
These were not merely overwhelming to the eye but caused automatic proof procedures to bog down: their execution time could rise exponentially.

The Isar proof language, introduced in the late 1990s, addressed these concerns by allowing proofs to be structured into nested scopes.
Local goals were proved from local assumptions, which were written out explicitly. 
Block structure is well understood in computer science, but here the ability to make declarations locally had to be retrofitted into Isabelle.
This led to the idea of local contexts to encapsulate the assumptions (and associated bound variables) 
specific to a particular goal, along with other information. During an ambitious proof, 
the user may still have several dozen assumptions available, 
but these are now structured through the nesting of the contexts and accessible by name rather than in a single giant list.

Locales are a further structuring mechanism for expressing an extended series of proofs that rest on shared assumptions.
They are useful for developing abstract mathematics, such as group theory, which can then be applied to particular groups.

\subsection{Structured Proofs: the Isar Language}
\label{sec:Isar}

A distinctive feature of Isabelle is its Isar language of structured
proofs \cite{wenzel-isabelle/isar}; the acronym stands for
\emph{Intelligible semi-automated reasoning}. In the original LCF
paradigm, a proof could be arbitrary ML code, which in some later
systems was replaced by a command language for proofs (e.g. the Ltac
scripting language in Coq). The problem with these traditional
approaches is that somebody looking at a machine proof can have no
idea what is being proved at a given point: it is like playing
blindfold chess. In contrast, an Isar proof is a hierarchical
structure containing explicit statements of assumptions and
conclusions, with an indication of the use of local facts.  Isar also
provides some mechanisms to avoid redundancy: it allows the proof
author to achieve a good balance of readability versus
maintainability, such that small changes to definitions and theorem
statements should lead to reasonably small changes to proofs.

Here is a tiny Isar proof that implicitly uses some derived rules for
logical connectives taken from the library (similar to the classical
reasoner from Section~\ref{sec:classical}):

\begin{isabelle}
\ \ \isacommand{have}\isamarkupfalse%
\ {\isachardoublequoteopen}A\ {\isasymand}\ B\ {\isasymlongrightarrow}\ B\ {\isasymand}\ A{\isachardoublequoteclose}\isanewline
\ \ \isacommand{proof}\isamarkupfalse%
\isanewline
\ \ \ \ \isacommand{assume}\isamarkupfalse%
\ {\isacharasterisk}{\isacharcolon}\ {\isachardoublequoteopen}A\ {\isasymand}\ B{\isachardoublequoteclose}\isanewline
\ \ \ \ \isacommand{show}\isamarkupfalse%
\ {\isachardoublequoteopen}B\ {\isasymand}\ A{\isachardoublequoteclose}\isanewline
\ \ \ \ \isacommand{proof}\isamarkupfalse%
\isanewline
\ \ \ \ \ \ \isacommand{from}\isamarkupfalse%
\ {\isacharasterisk}\ \isacommand{show}\isamarkupfalse%
\ B\ \isacommand{{\isachardot}{\isachardot}}\isamarkupfalse%
\isanewline
\ \ \ \ \ \ \isacommand{from}\isamarkupfalse%
\ {\isacharasterisk}\ \isacommand{show}\isamarkupfalse%
\ A\ \isacommand{{\isachardot}{\isachardot}}\isamarkupfalse%
\isanewline
\ \ \ \ \isacommand{qed}\isamarkupfalse%
\isanewline
\ \ \isacommand{qed}\isamarkupfalse%
\end{isabelle}
Here we prove a formula, \isa{A\ {\isasymand}\ B\ {\isasymlongrightarrow}\ B\ {\isasymand}\ A}.
We first assume \isa{A\ {\isasymand}\ B}, giving it the label~\isa{\isacharasterisk}.
Then we show \isa{B\ {\isasymand}\ A}, treating \isa{B} and \isa{A} separately.
Those subproofs refer to the assumption via its label.

The Isar approach scales from a few primitive inferences, as above, to
large proof developments involving heavy automated reasoning tools,
allowing the user to control the extent of proof automation.  The proof engine is
able to check well-structured Isar proofs more efficiently than
traditional tactic scripts: the hierarchical structure helps to keep
internal goals concise, without the intrusion of redundant assumptions or
unused lemmas. Moreover, the compositionality of Isar proofs, together
with the proof irrelevance of the Isabelle framework, allows independent
checking of sub-proofs---even with parallel checking on multiple
cores enabled by default, due to Matthews and Wenzel
\cite{Matthews-Wenzel:2010,Wenzel:2013:ITP}.

\smallskip The Isar proof language reuses some ideas from Mizar, a
legendary software tool for doing mathematics by machine
\cite{Grabowski2015}. But the many impressive facilities of the Mizar
language are intertwined with its unusual set theoretic formalism,
making the key ideas difficult to extract, especially given Mizar's
notorious lack of documentation and closed sources. In contrast,
Isar's generic principles of proof---fixing local variables and
making local assumptions---are identified with the corresponding
elements of Isabelle's logical framework. Thus it works with any
object-logic that uses the builtin Natural Deduction paradigm of
Isabelle, e.g. HOL, FOL, ZF. Various derived Isar language elements
provide explicit proof structure for typical reasoning seen in
object-logics: existential elimination and case-splitting,
calculational chains of equalities and inequalities (including
substitution), structured induction etc. All of this is parameterised
by declarations in the library.

The Isar proof language is parametrised by user-defined Isar
\emph{proof methods}, which are the old idea of ML tactics fitted into the
richer structure of the Isar engine. Thus new reasoning patterns may
be added to a theory library without revisiting the design of the
proof language itself. For example, the Isabelle standard library
defines method \emph{rule} for declarative forward/backward chaining
explained in Section~\ref{sec:early}, and method \emph{auto} for a
combination of the simplifier and the classical reasoner explained in
Section~\ref{sec:classical}. Proof methods are either defined in ML
(as in LCF or HOL), or in the Eisbach language
\cite{Matichuk-Murray-Wenzel:2016:JAR} (similar to Ltac definitions in
Coq): Eisbach uses the source notation of existing proof methods to
define new ones via simple recursion and pattern matching.

\smallskip Despite its importance for the Isabelle ecosystem, the Isar
proof language has not been adopted by other interactive theorem
provers. There are no fundamental obstacles to doing so,
but it takes some effort to do properly.  Some isolated aspects of
Isar have made it into the SSReflect language for Coq
\cite{Gonthier-Mahboubi:2010}, notably the \texttt{have} keyword for
local claims within a proof. A bit more Isar syntax made it into the
Lean prover \cite{de-Moura-et-al:2015} with slightly different
meaning, though. Lean's \texttt{fix}, \texttt{assume}, \texttt{have},
\texttt{show}, \texttt{obtain} construct certain
$\lambda$-terms in a more elementary manner than the Isabelle/Isar proof
context export and goal refinement operations; e.g.\ see the treatment
of \textbf{fix}-\textbf{assume}-\textbf{show} in
\cite[\S2.2]{wenzel-isabelle/isar}. The experiments by Wiedijk towards
supporting structured proofs in HOL Light \cite{wiedijk-mizar-light} are not directly related to Isar. 
They are simply a family of HOL Light tactics that allow a proof script written in ML to have some similarity to a Mizar text. 
This style of working has not caught on in the HOL world.

Although some detailed aspects of Isar are specific to Isabelle---forward/backwards refinement via
higher-order unification (Section~\ref{sec:early}) and local proof
contexts (Section~\ref{sec:context})---the language design as a whole is generic.
It requires nothing from the underlying calculus but primitive notions of terms and types 
and logical notions equivalent to ``implies'' and ``for all''.
It's not too hard to envisage a common language to be shared among proof assistants, 
offering portability of proofs exactly as today's programming languages offer portability between different machine architectures.
The crucial missing ingredient is the ability to write assertions and prove subgoals 
while minimising explicit references to a particular calculus.

\subsection{Global Theory Context}\label{sec:theory}

The original LCF approach (Section~\ref{sec:lcf-and-hol}) declares an
abstract ML type \texttt{thm} of theorems, but the
background theory is implicit in the system state: 
during a session there is only one big
theory under development, which grows monotonically.
There are primitives to introduce new types, terms, definitions and axioms
which augment the theory, but without any way to
undo such a change. HOL provides some tricks to hide
unwanted constants via name-space manipulation, to help interactive
development.

In contrast, as early as 1986, Isabelle declared the abstract ML type
\texttt{theory} alongside the type of theorems. 
Type \texttt{theory} makes available the initial theory of the logical framework
and allows for its extension as an acyclic graph of application
theories: there are operations to extend and merge theories.
Multiple theories can coexist in a single Isabelle/ML session, but end-users
only work with a single theory document at a time.
This may import other theory documents---in foundational order from bottom
to top. The concrete syntax looks like this:

\begin{isabelle}
\ \ \isacommand{theory}\ Test\isanewline
\ \ \ \ \isakeyword{imports}\ Main\ {\isachardoublequoteopen}HOL-Library{\isachardot}Finite{\isacharunderscore}Map{\isachardoublequoteclose}\ {\isachardoublequoteopen}HOL-Library{\isachardot}Finite{\isacharunderscore}Lattice{\isachardoublequoteclose}\isanewline
\ \ \isakeyword{begin}\isanewline
\ \ \ \ \isacommand{definition}\isamarkupfalse%
\ {\isachardoublequoteopen}constant\ {\isacharequal}\ term{\isachardoublequoteclose}\isanewline
\ \ \ \ \isacommand{theorem}\ name{\isacharcolon}\ statement\ \ \isacommand{{\isasymproof}}\isanewline
\ \ \isakeyword{end}
\end{isabelle}
Here we declare a new theory, called \isa{Test}. It is built upon three other theories: \isa{Main}, 
\isa{Finite{\isacharunderscore}Map} and \isa{Finite{\isacharunderscore}Lattice}. The last two names are qualified with the Isabelle \textit{session} to which they belong: \isa{HOL-Library}. Qualified names eliminate the danger of name clashes between theories belonging to different sessions.
Between the \isakeyword{begin} and \isakeyword{end} brackets we can make definitions and prove theorems.

Proved results are formally \emph{certified} against their
original theory context, e.g.\ a theorem $\Theta \vdash \phi$ for
theory~$\Theta$, and are implicitly propagated to another theory
$\Theta' \vdash \phi$, provided that $\Theta'$ extends $\Theta$ by
construction. For efficiency, this theory relation is implemented via symbolic
\emph{stamps} that represent definitions, proofs, etc., extending a theory: actual
theory content is not compared. Stamp inclusion needs to be checked in every inference step:
its complexity is logarithmic in the number of theory extensions,
which can be many thousands in typical applications.
These mechanisms lie within the logical kernel.

\begin{figure}
  \centering
  \includegraphics[height=12cm]{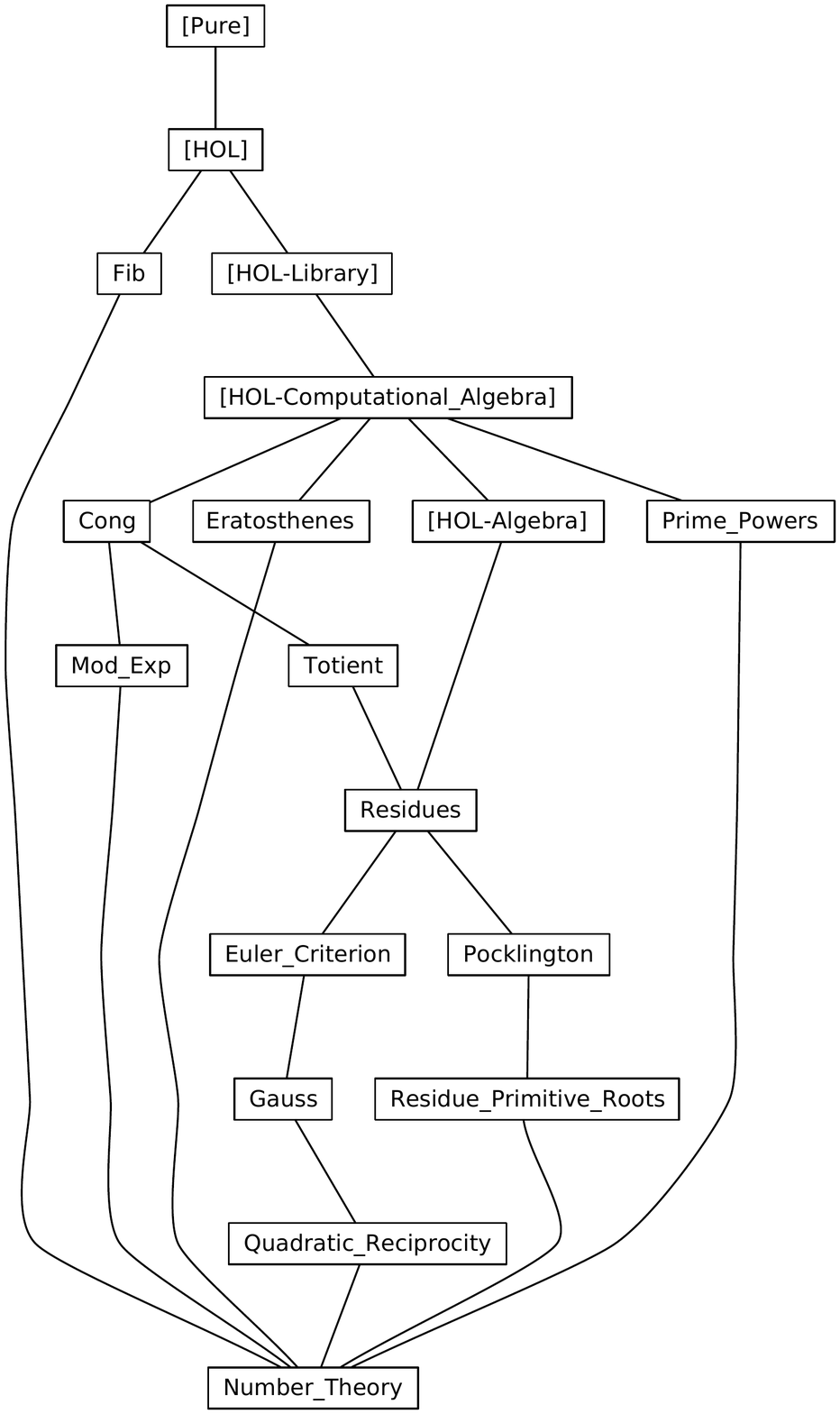}
  \caption{The theory dependencies of session \isa{HOL-Number\_Theory}} \label{fig:theory-deps}
\end{figure}

\smallskip In early versions of Isabelle as a logical framework,
theories coincided with object-logics plus some examples on top of
them: Isabelle/FOL, Isabelle/ZF, Isabelle/HOL etc. Derivations in
different branches of the theory graph could coexist in a single file without interference.
Today, theories are usually derived from the \isa{Main} entry-point of
Isabelle/HOL and built as a natural hierarchy according to the
structure of the application. It helps users to organise formal
definitions and proofs like consecutive chapters in a book; it also
helps the system for parallel checking of independent paths in the
theory graph. Figure~\ref{fig:theory-deps} shows the theory hierarchy of the Isabelle/HOL number theory development. 
At the top are sessions such as \isa{Pure} and \isa{HOL-Library}; 
in the middle are theories such as \isa{Totient}, all of which are imported into \isa{Number\_Theory}.

Isabelle theory operations are purely functional updates, making undo trivial: 
the earlier versions continue to exist and can be returned to. This also includes add-on content
like the ML environment or hints for proof tools: thus the theory
context provides a default set of parameters, according to the imports
from the library.

\smallskip In summary, Isabelle theories provide large-scale structure
to formalisation projects, with a built-in notion of monotonic
reasoning over an acyclic graph of theory nodes.

\subsection{Local Proof Context}\label{sec:context}

LCF and the HOL family lack an explicit notion of proof context.
There are usually some auxiliary structures to manage the current
proof state (maybe just a list of subgoals) for interactive theorem proving. However, proof tools
cannot refer to those; they merely see an isolated goal. Such a goal is essentially a sequent: 
a list of formulas (the assumptions) paired with another formula (the goal itself).
Proof assistants based on dependent type theories (like Coq) do have a
formal context $\Gamma$ that declares variables, but this again is essentially a sequent.
When applying HOL tactics it is frequently necessary to re-state the types of variables 
present in the statement of the theorem being proved, 
and one can even introduce two variables called \isa{x} with different types.
So at its most basic, we need a context to associate a type with each variable involved in the current proof. 
In fact, they do much more.

A genuine \emph{Local Proof Context} implemented by the ML type
\texttt{Proof.context} first appeared in Isabelle99, as
infrastructure for the then emerging Isar proof language
(Section~\ref{sec:Isar}). The idea is to support a block-structured
\emph{notepad} with recursive nesting of local declarations (e.g.\
type parameters, term parameters, assumptions) and local conclusions
that are generalized when leaving a nested block: there is a generic
\emph{export} operation to move results from a nested context into
the enclosing context: this usually turns context elements into the rule
structure of the Isabelle logical framework (the quantifier \isa{\isasymAnd} and
connective \isa{\isasymLongrightarrow}). In concrete syntax, this
looks as follows:

\begin{isabelle}
\ \ \isacommand{notepad}\isanewline
\ \ \isacommand{begin}\isanewline
\ \ \ \ \isacommand{{\isacharbraceleft}}\isanewline
\ \ \ \ \ \ \isacommand{fix}\ x\ y\ z\isanewline
\ \ \ \ \ \ \isacommand{assume}\ {\isachardoublequoteopen}A\ x{\isachardoublequoteclose}\ \isakeyword{and}\ {\isachardoublequoteopen}B\ y{\isachardoublequoteclose}\isanewline
\ \ \ \ \ \ \isacommand{have}\ {\isachardoublequoteopen}C\ x\ y\ z{\isachardoublequoteclose}\ \ \isacommand{{\isasymproof}}\isanewline
\ \ \ \ \isacommand{{\isacharbraceright}}\isanewline
\ \ \ \ \isacommand{note}\ {\isacartoucheopen}{\isasymAnd}x\ y\ z{\isachardot}\ A\ x\ {\isasymLongrightarrow}\ B\ y\ {\isasymLongrightarrow}\ C\ x\ y\ z{\isacartoucheclose}\isanewline
\ \ \isacommand{end}
\end{isabelle}

\noindent Here the final \isakeyword{note} recalls the result from the
preceding proof block (enclosed in the curly brackets). 
This block introduces three bound variables, assumed to be fixed 
and to satisfy \isa{A\ x} and \isa{B\ y}; from these we prove \isa{C\ x\ y\ z}. 
The effect of this block is to prove the formula shown in the note.

The \isakeyword{notepad} leaves nothing behind in
the theory context: it is merely an experiment. In contrast, a
\isakeyword{theorem} statement produces an initial proof context with
a pending claim: the theorem to be proved.
The subsequent proof body needs to solve that in the
context, potentially with further nesting of local contexts and
auxiliary claims. The final result extends the enclosing theory
(section \ref{sec:theory}) by the new fact.

Proof tools may access the \texttt{Proof.context} value at each point.
It contains the logical content (parameters, assumptions, proved
facts), local syntax, and hints to guide proof strategies.  Thus the
universal \texttt{Proof.context} replaces Isabelle's earlier tool-specific contexts
like \texttt{simpset} for the simplifier or \texttt{claset} for the
classical reasoner. This uniformity is important to combine tools:
e.g.\ the simplifier uses the context to extract its own
information (rewrite rules and auxiliary proof procedures); during the
simplification process it augments the context and passes it on to other
procedures, which in turn extract their own information from it (e.g.\
for classical reasoning).
Users can even extend the context with additional components to support 
proof procedures specific to their own applications. Contexts provide extensibility along with modularity.

\smallskip In summary, a proof context in Isabelle represents a local
situation derived from the enclosing theory. Nested contexts appear
and disappear; only exported results remain. For proof tools, the
context is a universal environment for storing tool-specific information.

%

\subsection{Structured Specifications: Locales}\label{sec:locales}

Isabelle locales provide an infrastructure for structured
specifications: definitions, statements and proofs of a theory may
depend on \emph{local parameters} (type and term variables) and
\emph{local premises} (hypotheses). The subsequent example specifies
partial orders axiomatically and defines a derived operation and
theorem in that specification context:

\begin{isabelle}
\ \ \isacommand{locale}\ partial{\isacharunderscore}order\ {\isacharequal}\isanewline
\ \ \ \ \isakeyword{fixes}\ le\ {\isacharcolon}{\isacharcolon}\ {\isachardoublequoteopen}{\isacharprime}a\ {\isasymRightarrow}\ {\isacharprime}a\ {\isasymRightarrow}\ bool{\isachardoublequoteclose}\ \ {\isacharparenleft}\isakeyword{infixl}\ {\isachardoublequoteopen}{\isasymsqsubseteq}{\isachardoublequoteclose}\ {\isadigit{5}}{\isadigit{0}}{\isacharparenright}\isanewline
\ \ \ \ \isakeyword{assumes}\ refl{\isacharcolon}\ {\isachardoublequoteopen}x\ {\isasymsqsubseteq}\ x{\isachardoublequoteclose}\isanewline
\ \ \ \ \ \ \isakeyword{and}\ trans{\isacharcolon}\ {\isachardoublequoteopen}x\ {\isasymsqsubseteq}\ y\ {\isasymLongrightarrow}\ y\ {\isasymsqsubseteq}\ z\ {\isasymLongrightarrow}\ x\ {\isasymsqsubseteq}\ z{\isachardoublequoteclose}\isanewline
\ \ \ \ \ \ \isakeyword{and}\ antisym{\isacharcolon}\ {\isachardoublequoteopen}x\ {\isasymsqsubseteq}\ y\ {\isasymLongrightarrow}\ y\ {\isasymsqsubseteq}\ x\ {\isasymLongrightarrow}\ x\ {\isacharequal}\ y{\isachardoublequoteclose}
\end{isabelle}
This introduces a new locale, called \isa{partial{\isacharunderscore}order}. The locale declares \isa{le}, effectively a constant to which we attach infix syntax. The locale asserts what are effectively axioms. But in reality, a locale abbreviates a predicate taking in this case a single argument, asserting that it satisfies the given assumptions. The point is to allow this small specification to be imported simply by quoting the name \isa{partial{\isacharunderscore}order}.

\begin{isabelle}
\ \ \isacommand{definition}\ {\isacharparenleft}\isakeyword{in}\ partial{\isacharunderscore}order{\isacharparenright}\isanewline
\ \ \ \ less\ {\isacharcolon}{\isacharcolon}\ {\isachardoublequoteopen}{\isacharprime}a\ {\isasymRightarrow}\ {\isacharprime}a\ {\isasymRightarrow}\ bool{\isachardoublequoteclose}\ \ {\isacharparenleft}\isakeyword{infixl}\ {\isachardoublequoteopen}{\isasymsqsubset}{\isachardoublequoteclose}\ {\isadigit{5}}{\isadigit{0}}{\isacharparenright}\isanewline
\ \ \ \ \isakeyword{where}\ {\isachardoublequoteopen}x\ {\isasymsqsubset}\ y\ {\isasymlongleftrightarrow}\ x\ {\isasymsqsubseteq}\ y\ {\isasymand}\ x\ {\isasymnoteq}\ y{\isachardoublequoteclose}\isanewline
\isanewline
\ \ \isacommand{theorem}\isamarkupfalse%
\ {\isacharparenleft}\isakeyword{in}\ partial{\isacharunderscore}order{\isacharparenright}\isanewline
\ \ \ \ less{\isacharunderscore}le{\isacharunderscore}trans{\isacharcolon}\ {\isachardoublequoteopen}x\ {\isasymsqsubset}\ y\ {\isasymLongrightarrow}\ y\ {\isasymsqsubseteq}\ z\ {\isasymLongrightarrow}\ x\ {\isasymsqsubset}\ z{\isachardoublequoteclose}\ \ \isacommand{{\isasymproof}}
\end{isabelle}
Now we can make definitions and prove theorems with respect to this locale, its constants and assumptions visible within these declarations.

Locales may be combined via \emph{locale expressions}, to rename or
instantiate parameters and merge contexts. Locale
\emph{interpretation} imports a given instance of a locale expression
into an application context: after proving the locale assumptions as
theorems, all conclusions of the locale context become available as
facts. Here is an example that instantiates abstract partial orders to
natural numbers:

\begin{isabelle}
\ \ \isacommand{interpretation}\ nat{\isacharcolon}\ partial{\isacharunderscore}order\ {\isachardoublequoteopen}{\isacharparenleft}{\isasymle}{\isacharparenright}\ {\isacharcolon}{\isacharcolon}\ nat\ {\isasymRightarrow}\ nat\ {\isasymRightarrow}\ bool{\isachardoublequoteclose}\isanewline
\ \ \ \ \isakeyword{rewrites}\ {\isachardoublequoteopen}nat{\isachardot}less\ {\isasymequiv}\ {\isacharparenleft}{\isacharless}{\isacharparenright}\ {\isacharcolon}{\isacharcolon}\ nat\ {\isasymRightarrow}\ nat\ {\isasymRightarrow}\ bool{\isachardoublequoteclose}\ \ \isacommand{{\isasymproof}}
\end{isabelle}

\noindent Thus \isa{le} is instantiated as the standard order on
type \isa{nat}, and the derived operation \isa{less} is identified
with the corresponding strict order on \isa{nat} (imposing a proof
obligation that needs to be proven together with the other locale
axioms). Afterwards, all conclusions from the
\isa{partial{\isacharunderscore}order} context become available in
terms of these native signatures on \isa{nat}.

Locales are ubiquitious in Isabelle theory development today, but the
majority are actually defined as type classes (section
\ref{sec:type classes}). These concepts started out independently, but
were unified by Haftmann and Wenzel \cite{HaftmannW06}. This is
another application of the locale interpretation concepts, which are due to
Ballarin \cite{Ballarin06}.

There are many ways of structuring mathematical developments via locales. 
One is simply to package up a group of related assumptions that are needed in a long series of proofs: this avoids cluttering up the theorem statements with this common material. Moreover, locales are more general than type classes. A type can belong to a class in only one way, so duality arguments (reversing the direction of a partial order for example) are out of the question; locales however can be instantiated in multiple ways at the same time. And while the type class for groups imposes the group structure on entire types, a locale for groups can have the carrier of the group as an explicit component of the locale; it can therefore be an arbitrary set. This generality is necessary to develop algebra properly. Isabelle's \isa{HOL-Algebra} session develops a substantial amount of group theory using this approach.

Ballarin \cite{Ballarin:2014} provides a comprehensive overview
of the concepts, and a tutorial on locales is part of the standard Isabelle
documentation. Many people have participated in the development of
locales over almost 15 years. The original work by Kammüller, Wenzel,
and Paulson \cite{kammueller-locales} directly uses primitives of the
Isabelle framework; important abstractions on top of
proof contexts (section \ref{sec:context}) were introduced by
Haftmann and Wenzel \cite{Haftmann-Wenzel:2009}.

\smallskip 
LCF did not have any concept comparable to locales.
But the need for structuring mechanisms has been evident for a long time,
and a variety of ideas were tried in systems as diverse as AUTOMATH, Coq and HOL\@.
The original locale concept was inspired by experiments (since abandoned) that had been done in HOL 
using higher-order predicates.
Coq has a richer logic with a built-in notion of ``sections'': the original Isabelle locales from
1999 \cite{kammueller-locales} were inspired by that, but without
augmenting the logical framework of Isabelle. This design principle of
building higher concepts without extending the
existing logic was later adopted by Coq: its type classes are
built on top of existing concepts like records, predicates, and
implicit arguments.

\subsection{ML within the Logical Context}

LCF and HOL can be seen as being nothing but libraries of proof procedures written in ML\@.
The user can call those procedures via the ML toplevel
but does not have to prove theorems and could instead, say, calculate $\pi$ to 10,000 digits.
In this sense, using LCF or HOL is the same as working with any other subroutine library. 

This type of ML toplevel no longer exists in Isabelle. Instead,
ML has become a sub-language of the framework. Syntactically, ML
expressions (or whole modules) may appear within the theory and proof
language of Isabelle/Isar: for example, in the command
\isacommand{method\_setup} to define a new proof method. Semantically,
program snippets may depend on symbolic entities from the logical
context: types, terms, facts, etc. The ML compiler is invoked at
run-time within an augmented environment; it refers to logical
entities as well as its own environment (for ML types, values,
modules). The static result is an updated ML environment within the
context: thus it also conforms to the parallel evaluation model of
Isabelle/ML.

Specialised proof procedures can be implemented in this manner within
the normal theory document, using the regular Prover IDE (see
Section~\ref{sec:PIDE}).  This integration of programming with logic
works without augmenting the logic: like in original LCF, ML has
access to the implementation of the object-logic, but is not part of
the logical formalism.  This is in contrast to Coq, where users often
implement proof tools inside the logical language itself (with
correctness proofs), but genuine extensions in OCaml need to be
assembled outside the system as ``plugins''.

\section{Document-oriented Interaction: the Prover IDE} \label{sec:PIDE}

The now ubiquitous WIMP interface (windows, icons, menus, pointer) emerged in the late 1970s, around the same time as LCF\@. 
Many observers suggested that the tedium of theorem proving could be addressed by involving those new ideas. 
However, typical suggestions did not address the real difficulties.
A favourite was to let users point to terms that must be rewritten. 
But in most situations it is infinitely more effective to execute such steps automatically, driven by an algorithm.
Think of dragging 10,000 files one-by-one to a trash can icon when they could be deleted by a single command specifying a pattern.
Many suggestions were geared to the needs of novices rather than to the professionals who would be the main users.

Useful interfaces for theorem provers would not appear for 20 years.
What users really needed, it seems, was the ability to survey the situation around them: what has been proved, what remains to be proved, what theorems are available, where and how they were proved, etc.

\subsection{Prover Interfaces: The Early Days}

The original LCF proof assistant from 1979 used a line-oriented terminal or teletype. 
This model is known as a read-eval-print loop (or REPL):
the user types one command after another, reacting on output printed
by the prover.  When computer screens and multiple windows arrived,
there was often a split into two areas: the editor to work on a
growing ``proof script'' and the terminal with the REPL to update its
state, using manual copy-paste operations from the editor.

Around 1998, the highly influential Proof General interface for Emacs
was released for the first time \cite{Aspinall:TACAS:2000}, including
support for Coq and Isabelle. Here a refined model of copy-paste and
state-synchronization is baked into the editor (which is freely
programmable in Lisp): the user can move a frontier of already checked
text either \emph{forwards} (apply command) or \emph{backwards} (undo
command); only the unchecked part may be freely edited. Proof General
requires suitable undo operations of the prover, and for robustness it
is better to have a restricted command-language instead of arbitrary
ML. Both are missing in HOL (any version), which consequently still uses
the original prover REPL, with some support through Emacs.

Today, Coq remains as the main back-end for Proof General Emacs,
there is also a popular Proof General clone written in OCaml: CoqIde.
In contrast, Isabelle discontinued both the REPL and its Proof
General mode in 2014: interaction now works exclusively via the
document-oriented Prover IDE.

\subsection{The Isabelle Prover IDE (PIDE)}

From 2008, multi-threaded ML programming became routinely
available in Isabelle, for parallel processing of theories and proofs
within a single Poly/ML process. This posed some challenges to
the robustness and performance of the prover engine, addressed subsequently
by Matthews and Wenzel \cite{Wenzel:2013:ITP}.

Parallel processing is also in conflict with the traditional
interaction model: the REPL acts like a single focus of
single-threaded command application. In order to remove many built-in
assumptions of sequential evaluation from the interaction model and to
provide rich semantic information in proof authoring process, Wenzel
introduced the document-oriented Prover IDE (PIDE) approach
\cite{Wenzel:2011:CICM,Wenzel:2012:UITP-EPTCS,Wenzel:2014:ITP,Wenzel:2019:MKM}. 
The main principles of PIDE are as follows:
\begin{itemize}

\item The \textbf{prover} supports \emph{document edits} and
  \emph{markup reports} natively. Interaction works via protocol
  commands (like \texttt{Document.update}) that take regular prover
  commands as data (e.g.\ \isakeyword{definition},
  \isakeyword{theorem}). It has its own policies to process proof
  documents in parallel, according to the structure of the text.

\item The \textbf{editor} connects the physical world of
  \emph{editor input events} and \emph{GUI painting} to the
  mathematical document-model of the prover. There are pipelines to
  stream input and output events asynchronously, with explicit
  identification of document versions.

\item Add-on \textbf{tools} may participate in the ongoing document
  processing by conventional means, as isolated functions from input
  to output that are managed by PIDE. External tools merely need to
  ensure that interrupts work correctly: this is required when the
  user continues editing and old versions of the document are
  discontinued eventually.

\end{itemize}

Unlike Proof General, PIDE never locks the source text: edits by the
user may lead to instantaneous updates by the prover, or significant
delays for slow proof tools. Thanks to overall performance
improvements of Isabelle and its underlying Poly/ML implementation,
this ambitious interaction model works smoothly.

Isabelle/PIDE is delivered to end-users as a fully integrated desktop
application called Isabelle/jEdit: it is based on the Java-based text
editor jEdit (see \url{http://www.jedit.org}). This explains the
initial motivation to use the Java platform for the outwards facing
side of PIDE, which is implemented in Scala (on the JVM). Non-Java
editors (e.g.\ VSCode) may be connected to PIDE via an extra socket
connection that exchanges JSON records.
There is also a \emph{Headless PIDE} server with a similar protocol;
this allows PIDE to run under control of another program.
  
Generally speaking,
Isabelle/ML works best for pure applications of mathematical logic
inside the prover, but Isabelle/Scala allows us to connect to the
physical world: IDE front-ends, database engines, TCP services
etc. Such technologies are not available in the same quality in
Standard ML, nor even in OCaml (which underlies HOL Light and
Coq).


\section{The Archive of Formal Proofs} \label{sec:afp}

Proof libraries are of enormous importance to formal verification. They are analogous to software libraries, facilitating reuse and eliminating the need to construct everything from scratch.

The Isabelle distribution already comes with a basic collection of more than 700,000 lines (38 MB) of Isabelle/HOL theories. On top of it sits the \emph{Archive of Formal Proofs} (AFP, see \url{https://www.isa-afp.org}), a large online collection of proof developments contributed by the Isabelle community, all of them (as of this writing) for Isabelle/HOL\@. Each entry or \textit{article} is a collection of Isabelle theories. 
It is the sole shared library of the Isabelle community. The AFP was launched in 2004 and
at the time of writing contains 480 articles written by 322 authors.
These comprise 2,250,000 lines (152 MB) of Isar text proving 134,000 theorems and supporting lemmas. It covers both computer science and mathematics.

As with a scientific journal, there is a small editorial board and submissions are reviewed for proof style and relevance. The AFP is an online resource and therefore more dynamic than a normal scientific journal. Articles can and do evolve.
This conflicts with the purpose of archiving entries as they have been submitted and with the purpose of providing a stable interface to users. However, true preservation requires ensuring that entries continue to work as Isabelle itself evolves.
The AFP deals with this conflict as follows.
For each Isabelle release there is a corresponding AFP release.
There is a separate  development version of the AFP that is updated by Isabelle developers and AFP authors:
Isabelle developers maintain all entries to be up to date with the current Isabelle development version;
authors can update their articles monotonically by adding further material while ensuring that all entries
  that depend on theirs still work.

As Isabelle evolves, the self-imposed requirement to maintain all AFP articles in working order puts a burden on the Isabelle developers. 
At the same time, it acts as a reality check,
helping developers to evaluate the impact of their changes on the user community.
Isar's structured proofs assist the maintenance effort by localising the impact of changes. Sledgehammer is also invaluable when fixing unfamiliar proofs.

The model for the AFP is the \emph{Mizar Mathematical Library} \cite{MML,BancerekBGKMNP18} which was started in 1989. The statistics today are similar: 250 authors, 1300 articles, 60,000 theorems, 3,000,000 lines (97 MB) of text. Of course one has to take into account that the languages and logics are different and Mizar has less proof automation.

\section{Postscript: Synergy between Ideas} \label{sec:post}

We have looked at a wide variety of ideas: a proof kernel written in a functional language; a logical framework to support multiple formalisms; polymorphism and type classes; advanced forms of automation; a structured proof language; a unique Prover IDE\@. Having seen the ideas in isolation, it's worth looking at how they work in combination.

Sometimes one idea led to another straightforwardly. Unification and backtracking were included by design to support future automation. The low-level primitives of the logical framework (implication and universal quantification) provided the right foundation for the Isar language precisely because they were the most fundamental logical concepts.

In other cases, the synergy between ideas could not have been predicted. It's remarkable that the early decision to adopt a polymorphic functional programming language (ML) turned out to be crucial 40 years later: the Isabelle Prover IDE is intertwined with pervasive parallelism. The Isabelle code base was not purely functional, but it was close enough, and the few sections that were necessarily imperative could be isolated easily. This yields an impressive speed up in multicore environments, a machine architecture nobody could have expected in 1975.

Another example of synergy is between automation (notably Sledgehammer) and the Isar language. The latter allows us to write a  derivation as a chain of simple steps, which can be proved automatically by the former. If a link of this chain needs a different sort of proof, such as induction, a nested scope can be inserted on the spot. This technique is particularly helpful to beginners, who would otherwise have to learn a great many specialised proof methods for transforming one formula into another, as with other proof assistants. But even experts don't have to think so hard when writing proofs, which are also easy to read because the chain of steps is written out explicitly. Powerful automation means the proofs don't have to be too detailed.

The earliest design decisions, dating from Edinburgh LCF, still make sense 40 years on. Our choice of a polymorphic functional language, a minimal proof kernel and no stored proofs yields good performance with a minimum risk of soundness errors; contrast that with alternative choices in many other automated theorem provers. Our focus on pervasive automation and readability must be contrasted with the prevailing tendency for low level, ``write only'' proofs. Our varied ideas have produced a system that looks like a unified whole, despite being the product of many people's contributions%
\footnote{Included in each release is a file entitled CONTRIBUTORS, but it only goes back to 2005 and has many omissions.}
over several decades.
 
\paragraph{Acknowledgement}
We thank the referees, Jasmin Blanchette, Cliff Jones, Michael Norrish and Andrei Popescu for valuable comments on drafts of this paper.
The work reported above was funded by the British EPSRC, the German DFG and various European Union funding agencies.

\bibliographystyle{alpha}
\bibliography{ideas}

\end{document}